\documentclass[
 reprint,
 amsmath,amssymb,
 aps,
 pra
]{revtex4-2}

\usepackage{amsmath}
\usepackage{graphicx}% Include figure files
\usepackage{xcolor}
\usepackage{bm}
\usepackage{upgreek}

\begin{document}

\title{Synthesis of Modulated Dielectric Metasurfaces for Precise Antenna Beamforming}%

\author{Vasileios~G.~Ataloglou}
\email{vasilis.ataloglou@mail.utoronto.ca}
\author{George~V.~Eleftheriades}%
\affiliation{The Edward S. Rogers Sr. Department of Electrical and Computer Engineering, 
University of Toronto, Toronto, Canada}
\date{November 2022}%
\begin{abstract}
This paper presents an end-to-end design method for the synthesis of dielectric metasurfaces with modulated thickness that are able to realize desired radiation patterns in the far-field with high precision. The method relies on integral equations and the Method of Moments to calculate the induced surface and volumetric current densities throughout the metasurface upon a given incident illumination. By avoiding homogenization of the dielectric layer and calculating directly the scattered fields through the Method of Moments, the near-field and far-field response of the metasurface can be accurately predicted and optimized. In particular, a design example of a modulated dielectric metasurface, fed by a single embedded line source, is presented for the realization of a Chebyshev array pattern with $-20 \ \mathrm{dB}$ sidelobe level. Moreover, an externally-fed dielectric metasurface performing beam splitting of a normally-incident plane wave is designed, 3-D printed and measured, showing two reflected beams at $\pm 30^\circ$, high suppression of specular reflections ($>14 \ \mathrm{dB}$) and satisfactory agreement with the simulated and desired radiation patterns. The proposed design method for beamforming with dielectric metasurfaces can find application in higher frequencies where the use of copper may be problematic due to higher ohmic losses and fabrication challenges.
\end{abstract} 
\maketitle

\section{Introduction}

Metasurfaces have demonstrated significant potential in manipulating electromagnetic waves in ways beyond the ones observed in natural material interfaces. Typically, metasurfaces take the form of subwavelength polarizable particles that are arranged in an electromagnetically thin sheet \cite{Yu:Science2011,Holloway:TAP2012,Pfeiffer:PRL2013,Selvanayagam:Opt2013}. These particles excite secondary co-located electric and magnetic dipole moments which scatter secondary fields, as envisioned by the Huygens' principle. By engineering locally the response to the incident illumination, a number of useful functionalities have been observed with the use of metasurfaces, including perfect anomalous refraction \cite{Lavigne:TAP2018,Chen:PRB2018}, perfect anomalous reflection \cite{Radi:PRL2017,Wong:PRX2018,Rabinovich:PRB2019}, antenna beamforming \cite{Epstein:TAP2017Antennas,Raeker:PRL2019,Dorrah:AWPL2018,Ataloglou:AWPL2021}, and polarization control \cite{Niemi:TAP2013,Selvanayagam:TAP2014}. 

The subwavelength dimensions of the scatterers comprising the metasurface allow its homogenization and its mathematical modelling with effective macroscopic surface parameters. These have been expressed in the literature in different forms, such as the effective surface polarizabilities, the surface susceptibilities or the surface impedances/admittances \cite{Kuester:TAP2003,Achouri:TAP2015,Epstein:TAP2016}. Following the underlying homogenization procedure, the design usually consists of two distinct steps. First, the macroscopic parameters are determined based on the input illumination and the desired functionality. Secondly, the homogenized parameters are implemented by appropriately designing the scattering elements. 

Lately, a number of works have been presented where the macroscopic design relies on a set of integral equations that accurately predict the scattered fields from the metasurface structure \cite{Budhu:TAP2020,Xu:Access2022, Budhu:TAP2022}. By iteratively solving the integral equations through a Method of Moments (MoM) \cite{Harrington:MoMBook}, it is possible to optimize one or more homogenized impedance layers to achieve a desired electromagnetic response from the metasurface. This full-wave optimization method captures the mutual coupling between all the metasurface scatterers, allowing for precise control of the near-field or far-field region. On the contrary, previous methods that represent the metasurface as an infinitesimally thin boundary and rely on the implementation of each discretized cell sometimes suffer from coupling between adjacent elements. While vias have been proposed as baffles between cells to suppress coupling effects, this approach increases fabrication challenges and it is practically applicable only to the transverse-magnetic (TM) polarization \cite{Xu:TAP2019}.

At microwave frequencies, metasurfaces commonly consist of one or more copper layers etched on standard dielectric substrates. Although the resonant nature of metallic scatterers allows for compact metasurface structures, it is also associated with increased losses, sensitivity to fabrication imperfections and limited bandwidth. Moreover, at higher frequencies, etching tolerances (e.g., minimum trace size) can become a limiting factor and require very delicate etching procedures. In these cases, dielectric-based metasurfaces could lead to designs that are less lossy, more broadband and easier to fabricate. In particular, the fabrication of dielectric metasurfaces is greatly facilitated by the advancements of 3-D printing technology, as 3-D printing materials with relatively low losses (e.g., Acrylonitrile Butadiene Styrene - ABS) can be easily printed in arbitrary shapes at a low cost \cite{Pepino:AWPL2021,Poyanco:EuCAP2021,Cheng:TAP2022}. 

Dielectric metasurfaces have been utilized in the past to realize beams based on the holographic approach \cite{Minatti:APS2010,Minatti:TAP2011}. The structure, consisting of a grounded corrugated dielectric, supports a single surface-wave mode that gradually converts to leaky-wave radiation by applying a slowly-varying modulation in the equivalent homogenized surface impedance. The surface impedance, then, translates locally to the corresponding height of the dielectric based on a transmission-line model. Some control of the radiation characteristics can be achieved by optimizing locally the modulation coefficient, that controls the intensity of surface-wave to leaky-wave conversion \cite{Puggelli:LAPC2015}. On the other hand, a recent work treated the design of impenetrable dielectric metasurfaces by optimizing the induced auxiliary surface waves to render a single homogenized impedance layer passive and lossless \cite{Lee:PRB2021}. It has been shown that these passively-induced surface waves are necessary to redistribute the incident power and achieve full phase and amplitude control with a passive metasurface \cite{Epstein:PRL2016,Kwon:PRB2018,Ataloglou:APS2020}. However, in  \cite{Lee:PRB2021}, the implementation of the optimized impedance layer with a thickness-modulated dielectric layer resulted to sub-optimal performance (compared to the same design with copper-based metasurfaces) due to the unaccounted coupling effects. Finally, a sparse dielectric metasurface controlling the reflected Floquet-Bloch (F-B) modes has been proposed; yet, the approach can only be applied to periodic metasurfaces with a relatively small number of propagating F-B modes \cite{Sharma:APS2021}.

In this work, we focus on the design of thickness-modulated dielectric metasurfaces to realize desired radiation patterns in the far-field region. The design is based on the optimization of the shape of the metasurface using integral equations solved with the MoM. Notably, the optimization scheme avoids homogenization completely, and, therefore, the derivation of an approximate relation between a homogenized surface impedance and the local height of the metasurface is not necessary. Moreover, all mutual coupling effects are considered and there is no assumption of local periodicity when determining the heights of each segment. This allows realizing beams with high-accuracy and desired characteristics both with an internal and an external illumination of the metasurface. Finally, the method can easily be extended to design dielectric metasurfaces that handle multiple inputs and produce multiple corresponding outputs. All examples are verified by commercial full-wave simulation solvers and a prototype for beam splitting in reflection mode is 3-D printed and measured.

The rest of the paper is organized as follows. Section~\ref{sec:Geometry} presents the geometry and mathematical formulation to analyze the structure through a set of integral equations. The optimization scheme and applied constraints are also described. A design example of a dielectric metasurface with an embedded source is discussed in Sec.~\ref{sec:Internal}. Section~\ref{sec:External} deals with the case of external illumination and a beam splitter is designed for a normally incident plane wave. Moreover, the fabrication details and the measurements to experimentally verify the design are presented. Lastly, Sec.~\ref{sec:Conclusion} concludes the paper.

\section{Analysis and synthesis framework}
In this section, we present the general geometry of thickness-modulated metasurfaces and the rigorous mathematical analysis of such structures through integral equations combined with the MoM approach. Subsequently, we introduce the optimization scheme to synthesize dielectric metasurfaces based on a desired far-field radiation pattern.

\subsection{Geometry} \label{sec:Geometry}
The geometry under consideration consists of a thickness-modulated dielectric layer of relative permittivity $\varepsilon_{r,1}$, an optional fixed-thickness dielectric layer of relative permittivity $\varepsilon_{r,2}$ and an underlying ground plane, as shown in Fig.~\ref{fig:Fig1}. The fixed dielectric layer has a thickness of $h_s$ and it can serve as a low-loss standard substrate for an embedded source, if the metasurface is internally illuminated. The structure has a total length of $L$ along the $y$-dimension, while it is assumed to extend infinitely along the $x$-axis to simplify our analysis. In practice, such uniformity along the $x$-axis can be established, even if we truncate the metasurface at around $5 \lambda$ ($\lambda$ being the free-space wavelength), as will be shown in simulations and experiments. The modulated layer is divided in $N$ segments with varying heights $h_n (n=1,2,...,N)$. The purpose of this work is to optimize the heights $h_n$ to obtain the desired far-field radiation. As mentioned, the incident field can either be produced by an embedded source, or it can refer to an incident wave illuminating the metasurface externally. In the latter case, the fixed dielectric layer can be omitted and the modulated layer is placed directly above the ground plane.

\subsection{Analysis through Integral Equations} \label{subsec:Analysis}
\begin{figure}
\includegraphics[width=0.88\columnwidth]{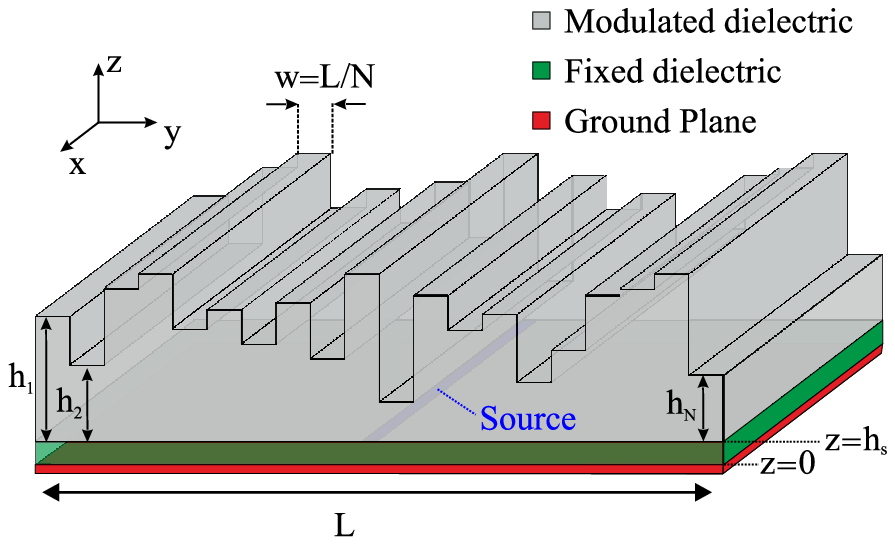}% Here is how to import EPS art
\caption{\label{fig:Fig1} Sketch of the geometry of the thickness-modulated dielectric metasurface.}
\end{figure}

To be able to optimize thickness-modulated dielectric metasurfaces, as the one shown in Fig.~\ref{fig:Fig1}, we first need to solve the analysis problem, i.e., to predict the scattering of the structure upon some kind of incident illumination. We base our analysis on a set of integral equations similar to \cite{Xu:Access2022}, with the main difference being that there is no homogenized impedance layer in the present work, but only a ground plane $C_g$ and a dielectric region $S_v$. We assume that the incident electromagnetic wave is transverse-electric (TE) polarized for simplicity $(\mathbf{E}_\mathrm{inc}=E_\mathrm{inc} \mathbf{\hat{x}})$. Together with the uniformity of the structure along the $x$-axis, it follows that the scattered fields would also be TE polarized and we can simply analyze a cross-section of the structure. Finally, a time-harmonic dependance $\mathrm{exp}\{+j\omega t\}$ is assumed for the electromagnetic quantities throughout the manuscript.

The incident field will induce a surface current density $J_g(\bm{\uprho})\mathbf{\hat{x}}$ along the ground plane $C_g$ and a polarization volumetric current density $J_v(\bm{\uprho}) \mathbf{\hat{x}}$ in the dielectric region $S_v$, where $\bm{\uprho}=y\mathbf{\hat{y}}+z\mathbf{\hat{z}}$ is the position vector in the $yz$-plane. In turn, the induced current densities produce secondary scattered fields $E_{\mathrm{sc},g}\mathbf{\hat{x}}$ and $E_{\mathrm{sc},v}\mathbf{\hat{x}}$, respectively. Since the dielectric is modelled with a polarization current, the corresponding scattered fields are calculated based on the $2$-D free-space Green's function as follows:
\begin{subequations} \label{eq:Scattered}
\begin{align}
E_{\mathrm{sc},g} \left(\bm{\uprho}\right) & = -\frac{k\eta}{4}\int\limits_{C_g} H^{(2)}_0\left(k\left|\bm{\uprho}-\bm{\uprho'}\right|\right)J_g(\bm{\uprho'})dy',\\
E_{\mathrm{sc},v} \left(\bm{\uprho}\right) & = -\frac{k\eta}{4}\iint\limits_{S_v} H^{(2)}_0\left(k\left|\bm{\uprho}-\bm{\uprho'}\right|\right)J_v(\bm{\uprho'})dy'dz',
\end{align}
\end{subequations}
where $H^{(2)}_0$  is the zero-order Hankel function of the second kind, $\bm{\uprho}$ and $\bm{\uprho'}$ are the observation and source position vectors, respectively, and $k=2 \pi/\lambda$ is the free-space wavenumber. From Eqs.~\eqref{eq:Scattered}, it is evident that calculating the scattered fields from the metasurface reduces to determining the unknown induced current densities $J_g, J_v$.

The total electric field $E_\mathrm{tot}$ is formed through the superposition of the incident field $E_\mathrm{inc}$ and the additional scattered fields $E_{\mathrm{sc},g}$ and $E_{\mathrm{sc},v}$. The total electric field should be zero at the ground plane (as it is tangential to it) and it should satisfy Ohm's law in the dielectric region:
\begin{align} \label{eq:OhmsLaw}
E_\mathrm{tot}\left(\bm{\uprho}\right)=E_\mathrm{inc}&\left(\bm{\uprho}\right)+E_{\mathrm{sc},g}\left(\bm{\uprho}\right)+E_{\mathrm{sc},v}\left(\bm{\uprho}\right) \nonumber\\ 
&=
\begin{cases}
 0 &, \mathrm{ on} \ C_g, \\
 \frac{1}{j\omega\left(\varepsilon_r\left(\bm{\uprho}\right)-1\right)\varepsilon_0} J_v (\rho) &,\mathrm{ in} \ S_v,
\end{cases}
\end{align}
where $\omega$ is the angular frequency of the wave, $\varepsilon_0=8.85 \times 10^{-12} \ \mathrm{F/m}$ is the free-space permittivity and $\varepsilon_r\left(\bm{\uprho}\right)$ is the position-dependant dielectric constant. In particular, the region $S_v$ refers to the whole rectangular region that can be potentially be filled up with the thickness-modulated dielectric. Based on the geometry of Fig.~\ref{fig:Fig1}, this means that $\varepsilon_r (\bm{\uprho})$ could take the values of $\varepsilon_{r,1}$, $\varepsilon_{r,2}$ or unity, depending on whether the observation point is in the modulated dielectric, the fixed dielectric or the air-filled region. In the latter case, the proportionality constant between the electric field and the volumetric current density becomes infinite, essentially indicating that there is no induced current in air.

\begin{figure}
\includegraphics[width=0.85\columnwidth]{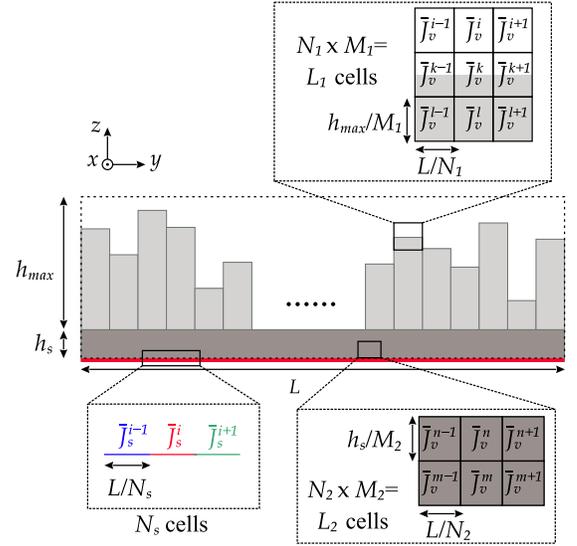}% Here is how to import EPS art
\caption{\label{fig:Fig2} Discretization grid for applying Method of Moments to determine the induced currents.}
\end{figure}

Equations \eqref{eq:Scattered} and \eqref{eq:OhmsLaw} form a set of integral equations with respect to the unknown current densities $J_s, J_v$ that we treat with the MoM. Specifically, the surface and volumetric current densities are expanded to non-overlapping $1$-D and $2$-D pulse basis function, respectively. The grid employed is illustrated in Fig.~\ref{fig:Fig2}. As observed, there are $N_s$ pulse basis functions on the ground plane $C_g$ corresponding to the surface current density. Similarly, there are $L_1=N_1 \times M_1$ pulse basis functions for the modulated-layer dielectric region and $L_2=N_2 \times M_2$ for the fixed-layer dielectric region. The total number of basis functions along $y$ in the modulated dielectric, denoted by $N_1$, is set to be an integer multiple of the number of segments $N$, so that no cell is shared between two consecutive segments. 

After expansion of the current densities to the basis functions, point matching is employed at the center of each $1$-D or $2$-D cell defined previously. This results in a linear system of equations of the following form:
\begin{align}\label{eqn:MoM_System}
\begin{bmatrix}
\mathbf{G}_{gg} &\mathbf{G}_{gv}\\
\mathbf{G}_{vg} &\mathbf{G}_{vv}-\mathbf{P}\\
\end{bmatrix}
\begin{bmatrix}
\bar{J}_g\\
\bar{J}_v\\
\end{bmatrix}=
- \begin{bmatrix}
\bar{E}_{\mathrm{inc},g}\\
\bar{E}_{\mathrm{inc},v}\\
\end{bmatrix},
\end{align}
where $\mathbf{G}_{ij}$ are matrices representing the self and mutual interactions between the different areas of the dielectric metasurface, $\bar{J}_i$ are the current amplitudes of the pulse pulse basis functions covering each cell and $\bar{E}_{\mathrm{inc},i}$ is the incident field values at the center of each cell. The expressions for the elements of the $\mathbf{G}_{ij}$ matrices are given in the Appendix~\ref{app:MoM_expressions}. Finally, $\mathbf{P}$ is a diagonal matrix of size $(L_1+L_2)\times (L_1+L_2)$ with the elements calculated as
\begin{align}
P[i][i] = \frac{1}{j\omega(\varepsilon_r^i-1)\varepsilon_0},
\end{align}
where $\varepsilon_r^i$ is the dielectric constant value at each cell. For a particular set of heights $h_n$, some cells may fall entirely within the dielectric region ($\varepsilon_r^i=\varepsilon_{r,1} \mbox{ or } \varepsilon_{r,2}$), some within the air region ($\varepsilon_r^i=1$) and some may intersect the interface between the modulated dielectric and air. For these intersecting cells, an averaged value based on the percentage of each region is assigned as the value $\varepsilon_r^i$ of the cell. This is important for facilitating the optimization of the heights $h_n$, since a smooth variation of the heights $h_n$ will result in a smooth variation of the averaged $\varepsilon_r^i$ in the cells of the interface and, as a result, an approximate, but smooth variation of the induced currents and the far-field radiation. On the contrary, assigning, for instance, the dielectric constant value of the middle of the cell could result to identical results for small variations of heights $h_n$ causing the optimization scheme to stop prematurely.

Solving the system in Eq.~\eqref{eqn:MoM_System} determines the induced currents $J_g$ and $J_v$ across the metasurface for a specified set of heights $h_n$. Then, the far-field electric field $E^{ff} (\theta)$ and the radiation intensity $U(\theta)$ at different angles $\theta$ can be computed as a superposition of the radiation from all the induced currents and from the incident source. The analytic expressions for this step are omitted, but they can be found in \cite{Xu:Access2022}.

\subsection{Synthesis through Optimization}
After treating the analysis problem in Sec.~\ref{subsec:Analysis}, the next step is to utilize it to form a synthesis optimization scheme that determines the heights $h_n$ of each segment in the modulated dielectric layer in order to achieve a desired electromagnetic functionality. The aim of this paper is to realize beams with desired far-field patterns $U_\mathrm{des}(\theta)$ in the $yz$-plane. Naturally, a cost function needs to quantify the difference between the desired radiation intensity and the one obtained for a dielectric metasurface with a set of heights $h_n$. Because the total power of the desired radiation intensity may differ from the available power from the source, the cost function takes a normalized form as follows:
\begin{align}
\label{eqn:cost_function}
F = \sum_{m=1}^{N_\theta} \left( \frac{U(\theta_m)}{\max_{\theta_m}\{U(\theta_m)\}}- \frac{U_\mathrm{des}(\theta_m)}{\max_{\theta_m}\{U_\mathrm{des}(\theta_m)\}} \right)^2,
\end{align} 
where $\theta_m$ is a set of $N_\theta$ equally spaced angles across the upper half space ($-\pi/2 \leq \theta_m \leq \pi/2$). It is also noted that radiation intensities $U(\theta_m)$ and $U_\mathrm{des}(\theta_m)$ are given in linear scale.

To minimize the cost function $F$ in Eq.~\eqref{eqn:cost_function} different optimization methods may be used. In this work, we utilize the built-in genetic algorithm of MATLAB, followed by gradient descent optimization. In general, a global optimization algorithm as the first step explores better the multivariable space of heights $h_n$ compared to the gradient descent that heavily depends on the initial point. On the other hand, gradient descent is utilized as a second step of the overall optimization, as it quickly converges to the (local) minimum around the initial solution. The pseudo-random nature of the seed in the genetic algorithm results in different solutions with every run of the optimization. Therefore, for challenging design cases, the combined optimization scheme can be ran multiple times and the best solution (the one corresponding to the minimum $F$ in Eq.~\eqref{eqn:cost_function}) is recorded.

Constraints can be introduced to converge to solutions that are practically realizable and have high performance metrics. In particular, the heights $h_n$ are constrained in the interval $[h_\mathrm{min},h_\mathrm{max}]$ in order for the metasurface to have structural support while maintaining compactness. Moreover, by allowing the dielectric constant to be complex, the power losses and the total radiated power can be predicted during optimization, as:
\begin{align}
P_\mathrm{loss}&=-\frac{\omega \varepsilon_0}{2} \iiint_{S_v} \mathrm{Im} \{\varepsilon_r \left(\bm{\uprho}\right) \} \left|E_\mathrm{tot}\left(\bm{\uprho}\right)\right|^2 dV, \\
P_\mathrm{rad}&=\int_{-\pi}^{\pi} U(\theta) d\theta.
\end{align}
Therefore, a constraint on the power efficiency $\eta=1-P_\mathrm{loss}/P_\mathrm{rad}$ can be placed fostering the convergence to solutions that exhibit less losses compared to the ones produced from unconstrained optimization, while they realize nearly identical radiation patterns. Typically, this means that lower-amplitude surface waves are excited to the vicinity of the metasurface in order to redistribute the incident power.

Finally, it is worth mentioning that the speed of our optimization scheme depends heavily on the inversion of the matrix in Eq.~\eqref{eqn:MoM_System}. To accelerate this process, we use the Kron reduction technique, as outlined in \cite{Xu:Access2022}. Based on this technique we can rearrange the equations in \eqref{eqn:MoM_System} to form a system of size $L_1 \times L_1$, that corresponds to the portion of the matrix that potentially changes at each iteration, as it depends on the cells of the thickness-modulated layer. Additionally, the rows and lines of the matrix corresponding to the air-filled cells can be discarded at each iteration, because there is no induced current in these cells, reducing the size of the inverted matrix further to only a portion of $L_1 \times L_1$.

\section{Internally-fed dielectric Metasurface} \label{sec:Internal}

In this section, we present the design of thickness-modulated dielectric metasurface illuminated by an embedded source. The structure has the form illustrated in Fig.~\ref{fig:Fig1} with a total length of $L=8 \lambda$ in the operating frequency of $10 \ \mathrm{GHz}$. The modulated dielectric is divided in $N=60$ segments with minimum height $h_\mathrm{min}=1 \ \mathrm{mm}$ and maximum height $h_\mathrm{max}=15 \ \mathrm{mm}$. The fixed-thickness layer has a height of $h_s=1.52 \ \mathrm{mm}$ and supports a line source located at $\{y_s,z_s\}=\{0,1.52 \ \mathrm{mm} \}$. We assume realistic dielectric constants of $\varepsilon_{r,1}= 3(1-j0.0045)$ for the fixed layer (close to the properties of ABS or other machinable dielectric materials) and $\varepsilon_{r,2}=3(1-j0.001)$ for the fixed layer (as provided, for example, for Rogers RO3003). The aim is to realize a Chebyshev radiation pattern with a sidelobe level (SLL) of $-20 \ \mathrm{dB}$. To determine the pattern $U_\mathrm{des}$, we first assume an array of $16$ virtual current line sources placed along the $y$-axis at a distance $\lambda/2$ between them and with currents $I_n$ calculated according to a Chebyshev antenna array \cite{Balanis:AntennaBook}. Then, the radiation pattern is simply the array factor of this linear non-uniform array.

\begin{figure}
\includegraphics[width=0.95\columnwidth]{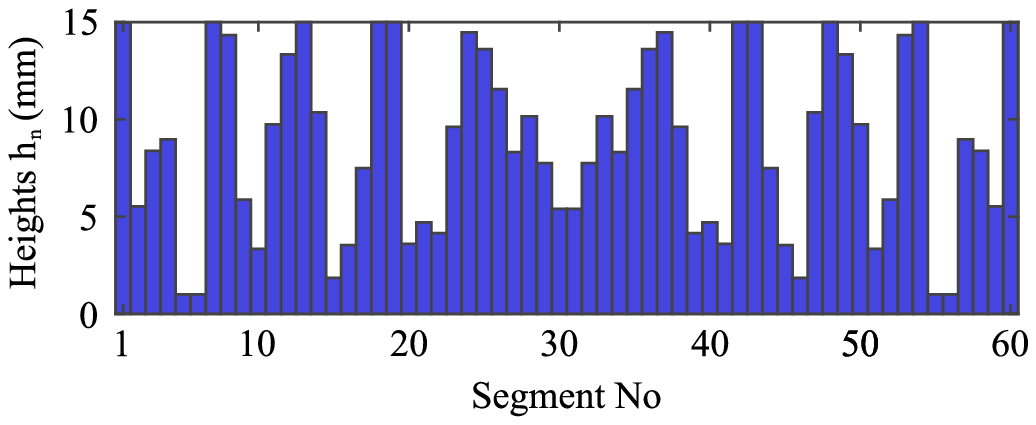}% Here is how to import EPS art
\caption{\label{fig:Fig3} Optimized heights $h_n$ for the internally-fed Chebyshev pattern dielectric metasurface antenna.}
\end{figure}

For the discretization grid required for the MoM framework, we use $N_s=600$ basis functions for the ground plane, $L_1=3600$ for the modulated thickness layer ($N_1=120$, $M_1=30$) and $L_2=1800$ for the fixed layer ($N_2=600$,$M_2=3$). Moreover, the far-field mismatch is minimized at a total of $N_\theta=361$ angles at the interval $[-\pi/2,\pi/2]$. Given that the incident field and the desired radiation pattern are symmetric along the $y$-axis, we also enforce this symmetry in the heights of the segments $h_n$, essentially reducing the optimization variables in half. Lastly, we constrain the power efficiency to $\eta \geq 88 \%$. The resulting optimized heights are plotted in Fig.~\ref{fig:Fig3}.

\begin{figure}
\includegraphics[width=0.9\columnwidth]{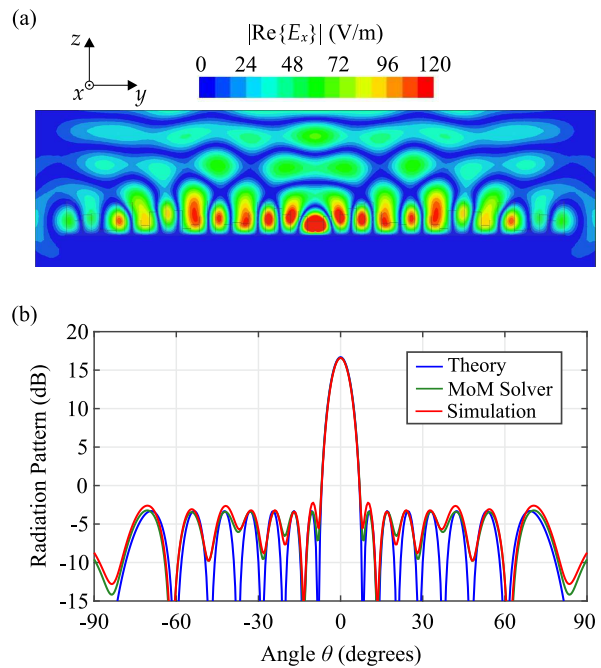}% Here is how to import EPS art
\caption{\label{fig:Fig4} Full-wave simulation results. (a) Field profile $|\mathrm{Re}\{E_x\}|$ in the $yz$-plane. (b) Radiation pattern obtained from the simulation (red curve) and the MoM solver (green curve) together with the desired Chebyshev pattern (blue curve).}
\end{figure}

We verify our framework by a full-wave simulation in Ansys High Frequency Structure Simulator (HFSS). As seen from the Fig.~\ref{fig:Fig4}(a), the incident power is redistributed along the dielectric metasurface and gradually leaks, forming a planar wavefront traveling towards $+\mathbf{\hat{z}}$ with a tapered amplitude corresponding to the Chebyshev pattern. The far-field radiation from the simulation is given in Fig.~\ref{fig:Fig4}(b), and it is compared with the results from our MoM solver and with the desired theoretical radiation pattern. Evidently, all of them match quite well in terms of the main beam, the null positions and the sidelobe levels. In particular, the obtained results from simulation differ by only $0.1 \ \mathrm{dB}$ from the desired pattern regarding the maximum directivity, while the SLL is kept at $-19 \ \mathrm{dB}$ or lower. The power efficiency coincided with the constraint of $88 \%$ for the solution of the MoM solver, and it is $89 \%$ in the HFSS simulation, limited primarily by the high-field values in the dielectric region close to the embedded source. Finally, the drop in the broadside directivity remains within $3$-dB compared to the nominal frequency for the range of $[9.65,10.6] \ \mathrm{GHz}$. This indicates a $9.5 \%$ relative bandwidth, which is higher compared to copper-based metasurfaces with similar functionality \cite{Xu:Access2022}.

\begin{figure}
\includegraphics[width=0.9\columnwidth]{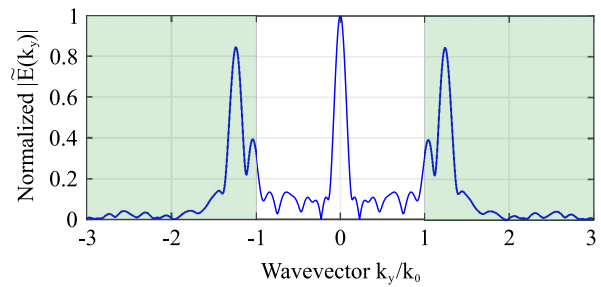}% Here is how to import EPS art
\caption{\label{fig:Fig5} Normalized spectral amplitude $|\tilde{E}(k_y)|$ of the electric field at $z=20 \ \mathrm{mm}$. The evanescent part $(k_y/k_0>1)$ is shaded.}
\end{figure}

It is worth examining the spectrum of the electric field in the vicinity of the dielectric metasurface. Specifically, we record the total electric field $E_\mathrm{tot}(y)$ at the plane $z=20 \ \mathrm{mm}$ (around $3.5 \ \mathrm{mm}$ above the maximum height of the structure) and calculate its spatial spectrum components. The spectral amplitude is depicted in Fig.~\ref{fig:Fig5}, where clearly a strong evanescent spectrum ($|k_y|/k_0>1$) is observed. These surface waves effectively redistribute the input power from the localized source towards the edges of the dielectric metasurface. Moreover, it is shown that an optimization targeting for a specific aperture field above the metasurface (in the near-field region) would be constraining for our far-field applications. This is because the near-field may contain strong evanescent waves that affect the electric field values and are not known \textit{a priori}, but they are still necessary to achieve the amplitude tapering and desired far-field radiation pattern.

\section{Externally-fed dielectric metasurface} \label{sec:External}
In this section, we discuss the design, fabrication and measurement of a thickness-modulated dielectric metasurface which is illuminated externally by a normally-impinging plane wave. Since there is no need for an embedded source, the fixed dielectric layer in Fig.~\ref{fig:Fig1} is omitted and the modulated dielectric is placed directly above the ground plane. The metasurface has a length of $L=7 \lambda$ at the frequency of $f=10 \ \mathrm{GHz}$ of the incident wave. This length is partitioned to $N=48$ segments, each with a width of $w \approx 4.38 \ \mathrm{mm}$ (or $\lambda/6.9$) and a height ranging between $h_\mathrm{min}=3 \ \mathrm{mm}$ and $h_\mathrm{max}=18 \ \mathrm{mm}$. The desired reflected fields are two superimposed plane waves radiating at $\pm 30^\circ$ with equal power in each one of them. To determine the desired far-field $U_\mathrm{des} (\theta)$, we assume two superimposed sets of virtual current line sources across the ground plane, each with uniform amplitude and linear phase:
\begin{align}
J_n= A_1 e^{-jk sin(30^\circ) y_n} + A_2 e^{+jk sin(30^\circ) y_n} e^{j \xi},
\end{align}
where  $\{y,z\}=\{y_n,z_n\}$ are the virtual sources positions, $A_1=A_2=1$ imposes the equal power-splitting and $\xi=90^\circ$ is selected as a phase detuning between the two beams for optimal sidelobe level. The far-field is calculated as the total radiation from these virtual sources. 

The dielectric constant of the modulated-thickness layer needs to be determined based on the available 3D-printable material; in our case, a standard ABS filament. For this purpose, we 3-D printed rectangular cuboids of solid ABS with their lateral dimensions fitting the cross-section of a WR-90 rectangular waveguide. By measuring the $S$-parameters of the waveguide with and without the samples, and using analytical expressions from the transmission through a lossy slab, we are able to deduce the values for the dielectric constant. Specifically, an average value of $\mathrm{Re} \{\varepsilon_r\}=2.5$ is extracted from the phase of $S_{21}$ at $10 \ \mathrm{GHz}$ and a loss tangent of $\mathrm{tan}(\delta)=0.01$ is estimated from the losses $A=1-|S_{11}|^2-|S_{22}|^2$. In any case, a small offset of the dielectric permittivity does not affect significantly the expected radiation pattern for our design. In the Supplemental Material, we include a number of supporting graphs showing the sensitivity of the optimized design to variations of the dielectric permittivity, the truncation of the metasurface along the $x$-axis and the wavefront of the illuminating wave \cite{Supplementary}.

\begin{figure}
\includegraphics[width=0.95\columnwidth]{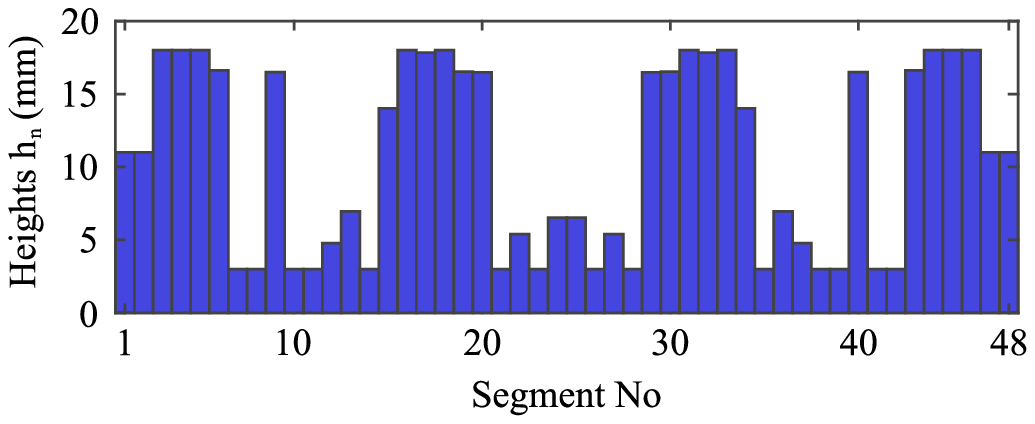}% Here is how to import EPS art
\caption{\label{fig:Fig6} Optimized heights $h_n$ for the externally-fed beam-splitting metasurface.}
\end{figure}

The optimization is performed with a discretization grid of $N_s=600$ basis functions for the ground plane and $L_1=2592$ basis functions for the dielectric area ($N_1=144, M_1=18$), while the radiation pattern is matched at $N_\theta=361$ angles at the interval $[-\pi/2,\pi/2]$. The power efficiency is constrained at $\eta \geq 0.90$ to avoid solutions that exhibit high field concentration within the lossy dielectric. The optimized heights after convergence are shown in Fig.~\ref{fig:Fig6}.

\begin{figure}
\includegraphics[width=0.85\columnwidth]{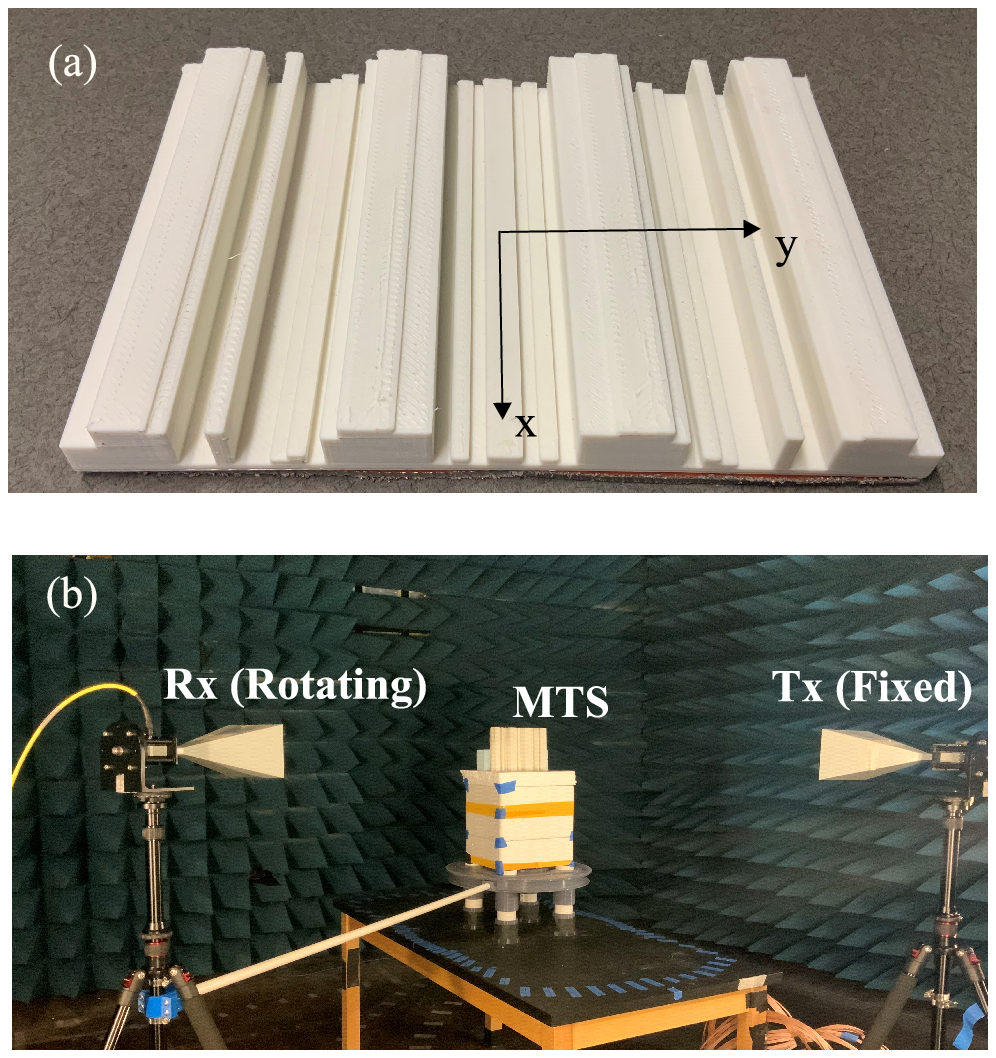}% Here is how to import EPS art
\caption{\label{fig:Fig7} (a) Printed prototype of the designed beam-splitting dielectric metasurface (MTS). (b) Measurement setup.}
\end{figure}

The optimized dielectric metasurface is 3-D printed with ABS in a Modix Big-60 3-D printer using a nozzle size of $0.8 \ \mathrm{mm}$ and $100 \%$ infill percentage for the dielectric. Moreover, the metasurface is truncated to $5 \lambda$ along the $x$-axis for fabrication, since this was proven sufficient in full-wave simulations to obtain a nearly-identical radiation pattern, when comparing with the infinite case. After fabrication, a copper-cladded substrate of the same dimensions is attached at the back of the dielectric to act as the underlying ground plane. A photo of the fabricated prototype is given in Fig.~\ref{fig:Fig7}(a).

The metasurface is measured in an anechoic chamber with the use of two identical X-band pyramidal horn antennas, as shown in Fig.~\ref{fig:Fig7}(b). Both horn antennas are placed at a distance of $\approx 150 \ \mathrm{cm}  \ (50 \lambda)$ from the stage holding the metasurface, ensuring that the transmitting antenna illuminates the metasurface with a planar wavefront and that the receiving antenna is recording values in the far-field region. The transmitting antenna is at a fixed location with its axis normal to the metasurface, while the receiving antenna rotates at a constant radius with the help of a moving tripod and a rotating pipe connecting the tripod to the stage.

\begin{figure}
\includegraphics[width=0.95\columnwidth]{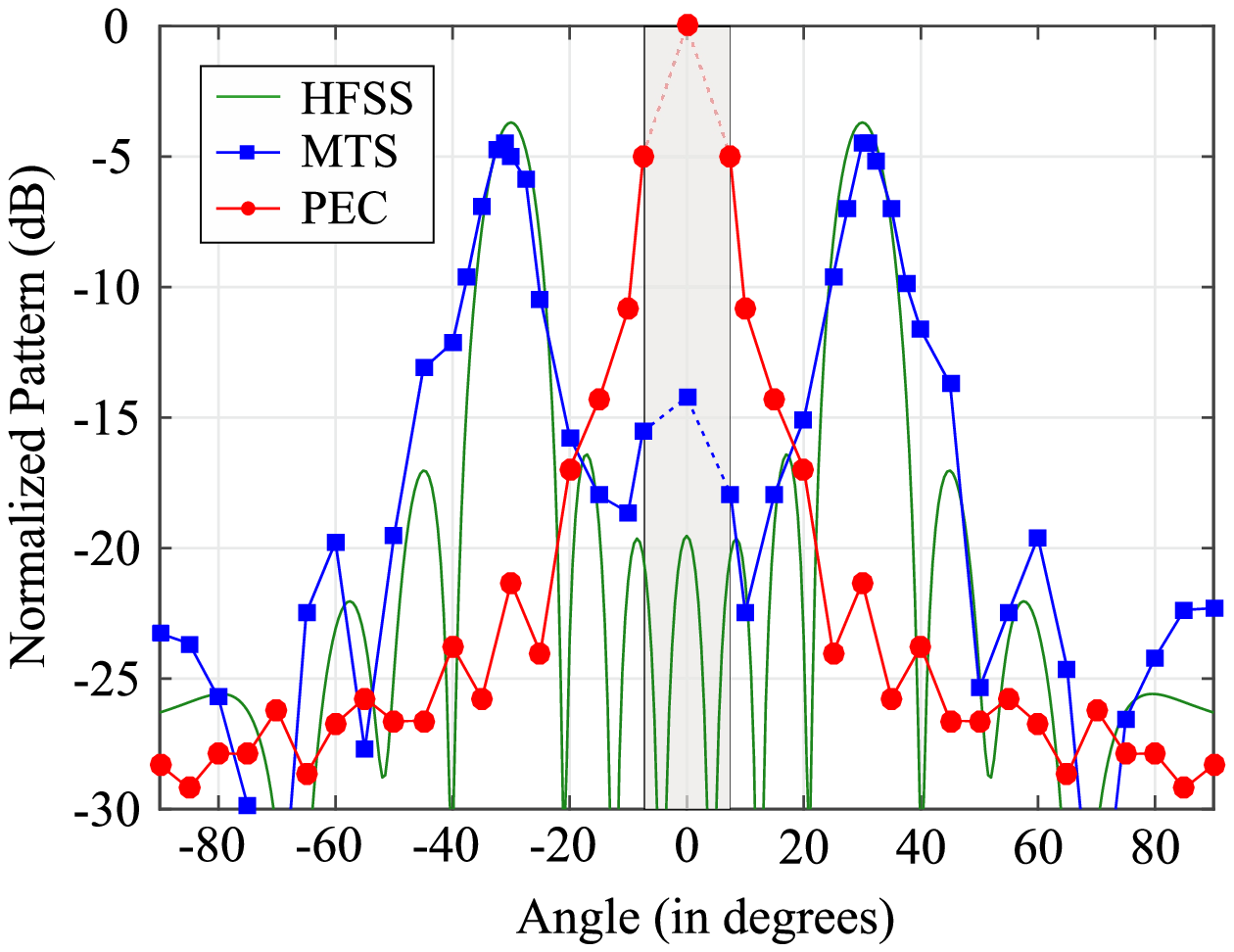}% Here is how to import EPS art
\caption{\label{fig:Fig8} Measured results for the beam-splitting metasurface. The normalized gain pattern (blue curve) is compared with the gain pattern from a PEC plate of the same dimensions (red curve) and the simulated results (green curve). The broadside results are recorded through $S_{11}$ instead of $S_{21}$ due to constraints of the measurement setup for $\theta < 7.5^\circ$.}
\end{figure}

The measured radiation pattern of the reflected fields, is given in Fig.~\ref{fig:Fig8}. Moreover, we have included the radiation pattern measurement of the copper base without the modulated dielectric (negative angles were mirrored for this measurement due to symmetry) and the simulated pattern from HFSS. Regarding the measurements, the receiving antenna physically collides with the transmitting antenna when their angular separation is less than $\Delta \theta= \pm 7.5^\circ$. However, since the two antennas have the same gain and are placed at the same distance from the stage, we use the value of $S_{11}$ for the measurement directly at broadside (instead of the $S_{21}$ which is used for all other angles). We also normalize all measured values with respect to the broadside reflections of the bare copper plate. Finally, the simulation pattern is also normalized with respect to the theoretical value of the radiation pattern at $\theta=0$ from a perfect conducting plate of the same size illuminated by a normally-incident plane wave.

As observed in Fig.~\ref{fig:Fig8}, two beams with the maximum intensity at $-31^\circ$ and $+30^\circ$ appear at the measured radiation pattern. Moreover, the specular reflections ($\theta=0^\circ$) are suppressed by more than $14 \ \mathrm{dB}$, when comparing the thickness-modulated metasurface with the bare copper plate. Generally, the pattern matches quite well the one predicted from simulation, especially when considering fabrication imperfections and the resolution of our bi-static measurement setup. Specifically, for the fabrication there was a slight warping for part of the metasurface, rendering its base non-flat and creating a non-uniform small air gap between the dielectric and the copper plate. This could be addressed in future designs by optimizing further the temperature settings for the 3-D printer or by fabricating the dielectric structure in smaller pieces before attaching them to the copper base. 

The peaks of the measured pattern are only $0.8 \ \mathrm{dB}$ below the simulation results. The observed difference includes both the drop due to the non-ideal pattern and also the power losses, which already account for $0.4 \ \mathrm{dB}$ ($9.3 \%$) in the simulation, assuming the extracted loss tangent $\mathrm{tan}(\delta)=0.01$. This demonstrates that indeed the dielectric beam-splitter exhibits quite low losses, a metric cared for through the constraints in the optimization phase. Lastly, to quantify the bandwidth of our reflective beam-splitting metasurface, we record the measured gain for the peaks at $-31^\circ$ and $+30^\circ$. The drop compared to the value in $10 \ \mathrm{GHz}$ is less than $1 \ \mathrm{dB}$ for the frequency range of $[9.05,10.5] \ \mathrm{GHz}$, while the same applies for the range $[9.2,10.8] \ \mathrm{GHz}$ in the simulation results.

While, we presented here a relatively simple example of symmetric beam splitting to experimentally verify the proposed method, the design opportunities of our MoM framework are vast, and they can include metasurfaces that manipulate multiple inputs to produce multiple desired outputs. A description of optimizing for such a design case and the corresponding simulation results are included in the Supplemental Material \cite{Supplementary}.

\section{Conclusion} \label{sec:Conclusion}
In conclusion, we present a design framework for dielectric metasurfaces with varying thickness for realizing desired radiation patterns. The method avoids homogenization techniques employed in previously designed corrugated dielectric metasurfaces. On the contrary, it is based on integral equations which accurately predict the induced current upon some incident illumination. In the first presented example, the illumination is coming from an embedded source and the metasurface radiates waves with a Chebyshev radiation pattern in the far-field region. Full-wave simulations verified the good matching with the desired pattern in terms of the directivity and the sidelobe level of $-20 \ \mathrm{dB}$. Next, an example with an externally-illuminated dielectric metasurface was presented for beam-splitting in reflective mode. This design was verified by 3-D printing such a prototype and performing measurements with a bi-static measurement setup. The experimental results showed beam-splitting at the desired angles of $\pm 30^\circ$ with high suppression of specular reflections and low power losses. The proposed end-to-end design framework, resulting in dielectric metasurfaces that can be directly fabricated with 3-D printing technology, offers beamforming capabilities and it can be advantageous in cases that copper-based metasurfaces are lossy, costly or hard to fabricate. Such beamforming functionalities with the use of dielectric metasurfaces can be utilized in existing and emerging technologies, such as 6G communication links, antenna radomes and dielectric cloaking coatings among others.

\appendix
\section{Mathematical expressions for the Method of Moments solution} \label{app:MoM_expressions}
For the MoM solution, the surface and volume electric current densities in Eqs.~\eqref{eq:Scattered} are expanded to a set of pulse basis functions with constant amplitudes $\bar{J}_g [m]$ and $\bar{J}_v [m]$, respectively. The centers of the pulse basis functions for the ground plane are located at $\bm{\uprho}_{gn}=y_{gn} \mathbf{\hat{y}} + z_{gn} \mathbf{\hat{z}}$ and for the dielectric cells at $\bm{\uprho}_{vn}=y_{vn} \mathbf{\hat{y}} + z_{vn} \mathbf{\hat{z}}$, with the subscript $n$ extending up to the total number of the corresponding cells. Then, point matching is applied at the center of each 1-D or 2-D cell, which is equivalent to taking the inner product with a delta function to both sides of Eq.~\eqref{eq:OhmsLaw}, based on:
\begin{equation}
<f,\delta(\bm{\uprho}-\bm{\uprho}_{in})>=\int f \delta(y-y_{in}) \delta(z-z_{in}) dx dy,
\end{equation}
where $i=\{g,v\}$ refers to the ground plane and dielectric cells, respectively. Based on this, a system of equations is obtained that is written in matrix form in Eq.~\eqref{eqn:MoM_System}. 

Here, expressions for the elements of the $\mathbf{G}_{ij} \ (\{i,j\}=\{g,v\})$ matrices in Eq.~\eqref{eqn:MoM_System} are given. For the non-diagonal elements ($n\neq m$) of the matrix $\mathbf{G}_{gg}$ and for all elements of matrix $\mathbf{G}_{vg}$, using the midpoint rule (with a single interval), we have:
\begin{align}
\mathbf{G}_{gg}[n][m]=-\frac{k \eta \Delta_s}{4} H^{(2)}_0\left(k\left|\bm{\uprho}_{gn}-\bm{\uprho}_{gm}\right|\right) , \\
\mathbf{G}_{vg}[n][m]=-\frac{k \eta \Delta_s}{4} H^{(2)}_0\left(k\left|\bm{\uprho}_{vn}-\bm{\uprho}_{gm}\right|\right) ,
\end{align}
where $\bm{\uprho}_{im}$, $\bm{\uprho}_{in}$ are the position vectors corresponding to the discretizing pulse basis function (source points) and the point matching locations (observation points), respectively and $\Delta_s=L/N_s$ is the width of each discretized segment in the ground plane. The diagonal terms of the matrix $\mathbf{G}_{gg}$ are given based on a small argument approximation for the Hankel function (Eq.~(3.14),\cite{Harrington:MoMBook}):
\begin{align}
\mathbf{G}_{gg}[n][n]=-\frac{k \eta \Delta_s}{4} \left[1-j\frac{2}{\pi}\log\left(\frac{1.781k\Delta_s}{4e}\right)\right],
\end{align}
where $e \approx 2.718$ is the Euler's number.

For the matrices $\mathbf{G}_{gv}$ and $\mathbf{G}_{vv}$, where the source point refers to the 2-D dielectric cells, the approximation introduced in \cite{Richmond:TAP1965} is used, which treats the 2-D rectangular source cell as a circular cell of equal area. Using this approximation, the following expressions can be derived for the matrix elements:
\begin{align}
& \mathbf{G}_{gv}[n][m]= -\frac{\eta \pi r_0}{2} J_1(kr_0) H^{(2)}_0\left(k\left|\bm{\uprho}_{gn}-\bm{\uprho}_{vm}\right|\right), \label{eq:G_gv} \\
& \mathbf{G}_{vv}[n][m]=  \nonumber \\
& \begin{cases} 
-\frac{\eta \pi r_0}{2} J_1(kr_0) H^{(2)}_0\left(k\left|\bm{\uprho}_{vn}-\bm{\uprho}_{vm}\right|\right), & n \neq m,\\
-\frac{\eta}{2k}\left[\pi kr_0 H_1^{(2)}(kr_0)-2j\right], & n = m,
\end{cases}\label{eq:G_vv}
\end{align}
where $J_1$ is the first-order Bessel function, $H_1^{(2)}$ is the first-order Hankel function of the second kind, and $r_0= \sqrt{\Delta_{y} \Delta_{z}/ \pi}$ is the radius of a circle with area equal to the size of the 2-D rectangular dielectric cell with dimensions $\Delta_y \times \Delta_z$.

%\bibliography{HMS_Nov2022.bib}

%

\end{document}

\title{Supplementary material for: \\ \vspace{1em} Synthesis of Modulated Dielectric Metasurfaces for Precise Antenna Beamforming}%

\author{Vasileios~G.~Ataloglou}
\email{vasilis.ataloglou@mail.utoronto.ca}
\author{George~V.~Eleftheriades}%
\affiliation{The Edward S. Rogers Sr. Department of Electrical and Computer Engineering, \\ University of Toronto, Toronto, Canada}
\maketitle

\section{Beam-splitter sensitivity simulations}
In the main text, a Method of Moments framework was presented for the design of modulated dielectric metasurfaces that can realize desired radiation patterns in the far-field. The metasurface was assumed to extend infinitely along the $x$-axis in Fig.~1 of the main text, allowing a simple analysis of the cross-section of the structure based on 2-D free-space Green's functions. However, in practice, the structure has to be truncated at a finite length for fabrication and measurements. In this section of the Supplementary Material, we examine the effect of this truncation to the far-field radiation pattern for the beam-splitting example discussed in Sec.~IV of the main text. Moreover, we investigate how other factors, such as a variation of the extracted permittivity of the ABS material or the illumination of the metasurface may affect the nominal results. These parametric analyses are important, as they demonstrate the robustness of the design to factors that emerge during fabrication and measurement of such modulated dielectric metasurfaces.

First of all, we truncate the designed metasurface to a finite length of $L_x=\lambda, 3\lambda \mbox{ and } 5\lambda$ along the $x$-axis and perform full-wave simulations in Ansys HFSS. By recording the radiation pattern of the reflected wave in the $yz$-plane, we compare the result with the metasurface that extends infinitely. As observed in Fig.~\ref{fig:FigS1}, a finite truncation of $L_x=3\lambda$ is sufficient to get a response very close to the one of the infinite structure that was analysed and optimized through the Method of Moments. If the length becomes much smaller (case for $L_x= \lambda$), the endfire radiation $(\pm 90^\circ)$ starts to increase and the direction of the main beams shifts slightly towards broadside. From the above analysis, it is obvious that the fabricated prototype with $L_x=5\lambda$ will not be affected by the truncation along the $x$ dimension.

\begin{figure}
\includegraphics[width=0.5\columnwidth]{FigS1.eps}% Here is how to import EPS art
\caption{\label{fig:FigS1} Effect of the truncation (along $x$) of the dielectric metasurface to the radiation pattern.}
\end{figure}

The relative permittivity of the dielectric may also cause discrepancies compared to the performance of the nominal design. Specifically, the dielectric constant $\varepsilon_r=2.5(1-j0.01)$ was extracted by characterizing different samples that were fabricated with the ABS filament in a WR-90 waveguide. However, the use of another ABS filament or the 3-D printing procedure of the modulated dielectric metasurface may introduce some uncertainty in the actual dielectric constant. To quantify such issues, we have performed simulations with $\varepsilon_r'=2.3(1-j0.01)$ and $\varepsilon_r''=2.7(1-j0.01)$. While the main beams are barely affected (less that $0.1 \ \mathrm{dB}$ drop, it can be seen from Fig.~\ref{fig:FigS2} that the specular reflections gets increased by around $5 \ \mathrm{dB}$ and $2 \ \mathrm{dB}$, respectively. On the other hand, varying the loss tangent leaves the radiation pattern practically unaltered, but scales linearly the power losses of the metasurface.

\begin{figure}[h]
\includegraphics[width=0.5\columnwidth]{FigS2.eps}% Here is how to import EPS art
\caption{\label{fig:FigS2} Radiation pattern for different values of the real part of the relative permittivity $\mathrm{Re}\{\varepsilon_r\}$ characterizing the modulated dielectric layer. The nominal design is based on $\varepsilon_r=2.5(1-j0.01)$.}
\end{figure}

Lastly, we investigate the effect of the illumination used in the experiment compared to the theoretical plane-wave illumination. Although the transmitting horn antenna is placed around $50 \lambda$ away from the center of the dielectric metasurface, the wavefront would still have some curvature causing a phase discrepancy at the incident wave. As a simple examination, we assume a cylindrical wave with its center coinciding with the transmitting horn antenna location, $50\lambda$ away from the dielectric metasurface. Figure~\ref{fig:FigS3} shows that the radiation pattern is also insensitive to this change. However, it can be observed that the nulls at $\pm 40^\circ$ have become more subtle. This effect, together with other fabrication imperfections (e.g. warping) and the limited resolution of our bi-static measurement system, may be responsible for the absence of these particular dips in the measured radiation pattern, as presented in Fig.~8 of the main text.

\begin{figure}[h]
\includegraphics[width=0.5\columnwidth]{FigS3.eps}% Here is how to import EPS art
\caption{\label{fig:FigS3} Simulation results for plane wave illumination (black dotted) and cylindrical wave illumination with the source $50\lambda$ away from the metasurface (blue curve).}
\end{figure}

\section{Dual-input dual-output design of Modulated Dielectric Metasurfaces}

The design examples in Sec.~III and Sec.~IV of the main text are based on modulated dielectric metasurfaces that act on a single input illumination and produce a single radiation pattern in the far-field region. However, the method can easily be extended to multi-input multi-output dielectric metasurfaces, where a single metasurface handles multiple illuminations to produce multiple corresponding output beams. In the context of communications, this effectively means an increase of the total capacity of the system, as two or more channels can be simultaneously radiated to beams in different directions. The input channels can refer to different spatially-separated embedded sources or to waves impinging to the metasurface from different illuminating angles. In this section, we describe the modifications on the method to design for such a case; specifically, two incident plane waves from $\theta_{\mathrm{in}}^1=-22.5^\circ$ and $\theta_{\mathrm{in}}^2=+22.5^\circ$ are anomalously reflected to $\theta_{\mathrm{out}}^1=+45^\circ$ and $\theta_{\mathrm{out}}^2=0^\circ$, respectively.

In terms of the mathematical formulation, the same steps are followed to optimize the heights $h_n$ of the modulated dielectric layer. The $\mathbf{G}$ matrices in Eq.~(3) of the main text depend purely on the geometry and cell grid for the Method of Moments solution, while the $P$ matrix depends on the heights $h_n$ which are identical for both inputs. However, the constant matrix on the right side of Eq.~(3) is different for the two illuminating waves, since it contains the values of the incident wave at the center of each cell. Therefore, the induced currents $J_s^{(i)}$ and $J_v^{(i)}$ can be calculated for the two incident waves ($i=\{1,2\}$). Based on these sets of currents, the far-field radiation patterns $U_1(\theta)$ and $U_2(\theta)$ are calculated. The optimization procedure is similar with the cost function being modified in order to account for both inputs. Moreover, in this case we focus on maximizing the directivity obtained in the output desired outputs rather than matching the whole pattern. Thus, we define the cost function as follows:
\begin{align}
\label{eqn:cost_function_modified}
F = - \mathlarger{\mathlarger{\sum_{i=1}^2}}  \frac{U_i(\theta_\mathrm{out}^i)}{\displaystyle\int_{-\pi/2}^{\pi/2} U_i(\theta) d\theta}.
\end{align}
Minimizing $F$ ensures radiation patterns with high directivities at the desired directions $\theta_{\mathrm{out}}^1=+45^\circ$ and $\theta_{\mathrm{out}}^2=0^\circ$ for the two incident waves. We note that the extension to multiple inputs can be also applied to incident waves at different frequencies. However, in this case, the impedance and polarization matrices in Eq.~(3) of the main text would be different for the two inputs, requiring double the amount of matrix inversions and, thus, nearly double the optimization time.

The dual-input dual-output dielectric metasurface operates at $f=10 \ \mathrm{GHz}$ and has a total length of $L=7\lambda$. The fixed layer is omitted and a  modulated dielectric layer with dielectric constant $\varepsilon_r=2.5(1-j0.01)$ (similar to the properties of the characterized ABS) is assumed directly above the ground plane. The modulated dielectric is divided to $N=70$ segments with heights ranging between $h_\mathrm{min}=2 \ \mathrm{mm}$ and $h_\mathrm{max}=22 \ \mathrm{mm}$. The discretization grid comprises of $N_s=600$ 1-D cells for the ground plane and $L_1=3080$ 2-D cells for the dielectric area ($N_1=140$, $M_1=22$). The optimized heights and desired functionality are depicted in Fig.~\ref{fig:FigS4}.

\begin{figure}[h]
\includegraphics[width=0.55\columnwidth]{FigS4.eps}% Here is how to import EPS art
\caption{\label{fig:FigS4} Optimized heights $h_n$ for the dual-input dual-output dielectric metasurface. The incident angles and the desired reflected angles are shown with arrows.}
\end{figure}

Full-wave simulations are performed at Ansys HFSS to verify our design. The scattered field profiles $|\mathrm{Re} \{ E_\mathrm{scat}^i \}|$ for both inputs are given in Fig.~\ref{fig:FigS5}(a). The scattered fields below the metasurface cancel the incident waves, while the fields above the metasurface represent the reflected waves at the desired angles. The radiation patterns are plotted in Fig.~\ref{fig:FigS5}(b), together with the ones predicted by the Method of Moments. The directivity in the simulation is $14.23 \ \mathrm{dB}$ and $15.83 \ \mathrm{dB}$ for the beams at $\theta_{\mathrm{out}}^1=+45^\circ$ and $\theta_{\mathrm{out}}^2=0^\circ$, respectively. We define the illumination efficiency $\eta_\mathrm{il}$, as the obtained directivity divided by the theoretical directivity of an aperture with uniform amplitude and linear phase $\mathrm{exp}\{j k \mathrm{sin}(\theta_\mathrm{out}) y\}$. This metric reaches $\eta_\mathrm{il}^1=0.84$ and $\eta_\mathrm{il}^2=0.87$ for the two input cases, despite that the waves are handled by the same physical aperture. Lastly, the power efficiency, which is limited by the dielectric losses, is calculated at $\eta_1=82.1 \%$ and $\eta_2=80.6 \%$ for the two inputs.

\begin{figure}[h]
\includegraphics[width=0.9\columnwidth]{FigS5.eps}% Here is how to import EPS art
\caption{\label{fig:FigS5} Full-wave simulation results for the dual-input dual-output dielectric metasurface. (a) Field profile  of the scattered fields for the two input cases showing the simultaneous anomalous reflection of the two incident beams. (b) Radiation patterns for the two input cases through simulation (dashed lines) and the Method of Moments solver (solid lines).}
\end{figure}